\documentclass[12pt]{article}
\begin{document}
\title{\bf On the fine structure constant in the Machian universe}
\author{{\bf Merab Gogberashvili}\\
Andronikashvili Institute of Physics\\
6 Tamarashvili Street, Tbilisi 0177, Georgia\\
and\\
Javakhishvili State University\\
3 Chavchavadze Avenue, Tbilisi 0128, Georgia\\
{\sl E-mail: gogber@gmail.com}
\\
\\
{\bf Igor Kanatchikov}\\
National Quantum Information Centre of Gda\'nsk, \\
ul. W\l. Andersa 27, 81-824 Sopot, Poland \\
{\sl E-mail: kai@fuw.edu.pl}
\date{September 26, 2010}
}
\maketitle
\begin{abstract}
Within the Machian model of the universe the dark energy is identified with the Machian energy of collective gravitational interactions of all particles inside the horizon. It is shown that the fine structure constant can be expressed in terms of the observed radiative, baryon and dark energy densities of the universe and the densities of various matter components are interrelated via it.

\end{abstract}

\vskip 0.5cm


A Machian model of the universe has been put forward in \cite{Gog1,Gog2,Gog3}. A new, 'thermodynamic' formulation of Mach's principle \cite{Mach} has been introduced: the rest mass of a particle is a measure of its long-range collective interactions with the statistical ensemble of particles inside the cosmological horizon. The assumption that all particles in the universe are involved in local interactions avoids anisotropic effective masses which arose in earlier treatments of Mach's principle and have been ruled out by precision measurements \cite{Exp}. Besides, it effectively weakens the observed strength of gravity, thus allowing us to explain the hierarchy problem in particle physics \cite{Gog1}. A consequence of the assumption that all the particles in the universe form a unified statistical ensemble is the appearance of a fundamental cosmological reference frame. In spite of this explicitly anti-relativistic assertion, the model is  potentially compatible with the existing cosmological and gravitational theories in the low energy regime, in the sense that it imitates basic features of special and general relativity theories, such as the relativity principle, local Lorentz invariance, weak equivalence principle, and it also obeys the so-called signal locality, i.e. the matter is not allowed to propagate faster than light \cite{Gog2}. Besides, in \cite{Gog3} it was shown that the Machian universe appears to be compatible with the main features of quantum mechanics: both the Schr\"odinger equation and the Planck constant have been derived within the Machian approach.

In this paper we show that estimations based on the Machian model of the universe lead to the expression of the fine structure constant in terms of the observable cosmological parameters related to the energy content of the universe.

Our Machian model postulates a non-local Machian interaction which emanates from the collective gravitational interaction of all $N$ particles in the universe which form a world statistical ensemble. The non-locality extends up to the horizon scale, so that all particles inside the cosmological horizon are gravitationally entangled. It is assumed that the universal constant of the speed of light $c \approx 3 \times 10^{8} ~ m ~ s^{-1}$ originates in the collective gravitational potential $\Phi$ of the whole universe which acts on any particle of the world ensemble: 
\begin{equation} \label{Phi}
c^2 = - \Phi = \frac{2 M G}{R} ~.
\end{equation}
Here $M$ and $R \approx 3\times 10^{26} ~ m$ are the total mass and the radius of the universe, respectively, and $G \approx 7 \times 10^{-11} ~ m^3~ s^{-2}~ kg^{-1}$ is the Newton constant. Note, that $\Phi$ and $c$ are universal constants due to the cosmological principle, i.e. the isotropy and homogeneity of the universe at the horizon scale $R$. 

The relation (\ref{Phi}) allows us to express the Mach's principle which relates the origin of the rest energy, or inertia, of a particle to its interactions with the rest of $(N-1)$ particles in the universe:
\begin{equation} \label{mPhi}
E = -m\Phi = mc^2~.
\end{equation}
The latter formula takes into account the contribution of the collective Machian interactions between all particles as follows. Because each particle in the universe interacts with all other $(N-1)$ particles, the number of interacting pairs is $N(N-1)/2$. The mean separation in the uniformly distributed interacting pairs is $R$. Let us assume for simplicity, that all particles have an identical mass $m$. Hence, the total Machian energy consists of $N(N-1)/2$ terms of magnitude $\approx 2G m^2/R$, and the Machian energy of a single particle, eq. (\ref{mPhi}), is given by:
\begin{equation} \label{E}
E = \frac{N(N-1)}{2}~\frac {2Gm^2}{R} ~.
\end{equation}

From eq. (\ref{E}) it follows that the contribution of the collective Machian interactions to the total mass of the universe is equal to:
\begin{equation} \label{M=N2m}
M_{Mach} \approx \frac {N(N-1)}{2} m \approx \frac{N^2}{2} m ~.
\end{equation}
Correspondingly, the total mass of the universe is of the order of $\sim N^2m$ and not $\sim Nm$, as one would expect under the usual assumption of the additivity of mass.

Note, that (\ref{Phi}) is equivalent to the critical density condition in relativistic cosmology: 
\begin{equation} \label{rho}
\rho_{c} = \frac{3M}{4\pi R^3} = \frac{3H^2}{8\pi G} ~,
\end{equation}
which is a manifestation of the fact that the space-time is almost flat. Here, $H \sim c/R \approx 2 \times 10^{-18} ~s^{-1}$ is the Hubble constant. Using the experimental values of $c$, $G$ and $H$, from (\ref{rho}) or (\ref{Phi}) we obtain an estimation of the total mass of the universe:
\begin{equation} \label{M}
M \sim \frac{c^3}{2 GH} \approx 10^{53}~ kg~.
\end{equation}

Other important parameter of the Machian model is the elementary action of a typical particle from the world ensemble:
\begin{equation} \label{A}
A = -\int_{t_1}^{t_2} dt~ E \approx - m c^2 \Delta t \approx - 2\pi
\hbar ~,
\end{equation}
which is identified with the Planck constant $\hbar \approx 7 \times 10^{-34}~ m^2~ kg~ s^{-1}$ \cite{Gog3}. In this formula, $\Delta t = t_2 -t_1$ is the mean time between the non-local interactions of particles within the horizon $\sim R$, or the response (feedback) time of the universe to the particle motion. It is estimated as follows:
\begin{equation} \label{Deltat}
\Delta t \sim \frac{R}{Nc} \sim \frac{1}{NH}~.
\end{equation}
Note, that because of the non-local Machian interactions of all particles of the world ensemble found inside the horizon volume $\sim R^3$, the effective mean separation between particles is $R/N$, so that the response time $\Delta t$ is much shorter than e.g. the mean free motion time of particles with the mean separation $\sim R / N^{1/3}$ in a dilute gas.

Now, assuming $M_{Mach}\sim M$, from (\ref{M=N2m}), (\ref{A}) and (\ref{Deltat}) we can estimate the total action of the universe:
\begin{equation} \label{A_U}
A_U = - \frac{M c^2}{H} \sim \frac{N^3}{2} A ~.
\end{equation}
From the identification of the elementary action $A$ in (\ref{A}) with $- 2\pi\hbar$ we also obtain an estimation of the number of typical particles in the universe:
\begin{equation} \label{N}
N \sim \left(\frac{2 A_U}{A} \right)^{1/3} \approx
\left(\frac{Mc^2}{\pi \hbar H} \right)^{1/3} \approx 10^{40} ~.
\end{equation}
This number, which appears here as one of the parameters of the Machian model, is known as one of Dirac's 'large numbers' which point to the existence of a deep connection between the micro and macro physics \cite{large}.

Equations (\ref{M=N2m}), (\ref{M}) and the assumption $M_{Mach}\sim M$ yield also an estimation of the mass of a typical particle in our simplified Machian universe:
\begin{equation} \label{m}
m \approx {\frac{2 M_{Mach}}{N^2}} \approx 2 \times 10^{-27}~ kg \approx 1 ~ GeV~c^{-2}~,
\end{equation}
which appears to be of the order of magnitude of the proton mass. Alternatively, we could postulate that the typical mass of a particle forming the world ensemble is the mass of a typical stable heavy particle, i.e. the proton, and then use eq. (\ref{Deltat}) in order to obtain its elementary action given by (\ref{A}), which would coincide then with the Planck constant.

Let us stress here a difference between our model and the standard approach, where the mass of the universe (\ref{M}) is estimated as the sum of masses of $\sim 10^{80}$ protons. In the Machian model  the energy of  long-range interactions of all particles is taken into account, which is missing in the standard approach. As a consequence,  according to eq. (\ref{M=N2m}), only $\sim 10^{40}$ protons, eq. (\ref{N}), is needed in order to account for the correct value of the total mass of the universe, eq. (\ref{M}). For this reason, it is natural to identify the Machian energy of all particles with the dark energy of the universe, so that
\begin{equation} \label{Edark}
\Omega_\Lambda = \frac{M_{Mach}}{M} \approx \frac{N^2m}{2M} ~.
\end{equation}

Now, let us consider a little bit more realistic model universe which includes both neutral and charged particles. We assume that the universe as a whole is neutral, i.e. a half of charged particles carries positive charge $+e$ and the other half have negative charge $-e$. The number of charged particles can be roughly identified with the number of baryons in the universe $N_b < N$. Then an estimation of the Machian energy of the baryon component of matter yields: 
\begin{equation} \label{Eb}
E_{b|Mach} = (2N_b N - N_b^2)~\frac {Gm^2}{R} \approx 2 N_bN~\frac {Gm^2}{R}~.
\end{equation}
We can also expect that the ratio (\ref{Edark}) of the Machian and total energy is also valid for the corresponding baryon contributions:
\begin{equation} \label{E/E}
\frac{E_{b|Mach}}{E_{b|tot}} = \Omega_\Lambda~, 
\end{equation}
where $E_{b|tot}$ denotes the total energy of the baryon component of the universe. Then the observed baryon density in the universe can be written in the form:
\begin{equation} \label{Omegab}
\Omega_b \approx \frac{E_{b|tot}-E_{b|Mach}}{Mc^2} \approx \frac{E_{b|Mach}}{Mc^2}~\frac{(1-\Omega_\Lambda)}{\Omega_\Lambda}~.
\end{equation}

Further, let us estimate the electromagnetic energy of all $N_b$ charged particles (i.e. $N_b/2$ interacting pairs) in the model universe. The fact that electric charges have two polarities, while the mass is always positive, leads to basic differences. The universe as a whole is neutral and, in contrast to the Machian energy, the total electromagnetic, or radiative energy consists of $N_b/2$ additive terms, i.e. 
\begin{equation}\label{Er}
E_r \approx \frac {N_b}{2} \frac{k_ce^2}{R} = \frac {N_b}{2} \frac{\alpha \hbar c}{R}~,
\end{equation}
where $k_c$ is the Coulomb constant and $\alpha$ is the fine structure constant. 

Equations (\ref{Eb}), (\ref{E/E}), (\ref{Omegab}) and (\ref{Er}) yield for the ratio of the radiative and baryon densities in the universe:
\begin{equation} \label{omega/omega}
\frac {\Omega_r}{\Omega_b} \approx \frac{E_r}{E_{b|Mach}}~\frac{\Omega_\Lambda}{(1 -\Omega_\Lambda)} \approx \frac{\alpha \hbar c}{4 N G m^2}~\frac{\Omega_\Lambda}{(1-\Omega_\Lambda)}~,
\end{equation}
whence it follows:
\begin{equation}\label{alpha}
\alpha = \frac{(1-\Omega_\Lambda)}{\Omega_\Lambda}~\frac{\Omega_r}{\Omega_b}~ \frac{4N G}{c}~ \frac{m^2}{\hbar}~.
\end{equation}
From (\ref{Phi}), (\ref{A}) and (\ref{Deltat}) we find:
\begin{equation}
\frac{m^2}{\hbar} = \frac{2\pi c}{NG}~ \Omega_\Lambda~.
\end{equation}
Finally, by inserting this formula into (\ref{alpha}), we arrive at the expression for the fine structure constant in terms of the parameters $\Omega_\Lambda, \Omega_b$ and $\Omega_r$:
\begin{equation}\label{alpha-1} 
\alpha \approx 8\pi (1 -\Omega_\Lambda) \frac{\Omega_r}{\Omega_b} ~.
\end{equation}

From the contemporary observations we know that $\Omega_r /\Omega_b = 1.09 \pm 0.03 \times 10^{-3}$ and $ \Omega_\Lambda = 0.74 \pm 0.03$ \cite{La-Li}. Using those values in (\ref{alpha-1}), we obtain:
\begin{equation}
\alpha \approx 7.1 \pm 0.6 \times 10^{-3} ~,
\end{equation}
which is surprisingly close to the experimental value $\alpha \approx 7.297\times 10^{-3}$.

From (\ref{alpha-1}) we can also obtain a relationship between the dark energy and the baryon energy densities using the observed values of the fine structure constant $\alpha$ and the radiation energy density $\Omega_r = 4.8 \pm 0.04 \times 10^{-5}$ \cite{La-Li}:
\begin{equation}
\Omega_{\Lambda} \approx 1 - 6 \Omega_b~, 
\end{equation} 
which appears to be consistent with  current observations. Note that in the standard cosmological models the relation between the baryonic matter and dark energy densities is an arbitrary parameter determined from observations. Using the relationship between the densities of various matter species $\Omega_i$ and $\Omega_\Lambda$ in the flat universe: 
\begin{equation}
\Omega_\Lambda + \sum_i \Omega_i = 1~,
\end{equation}
we can exclude $\Omega_\Lambda$ from (\ref{alpha-1}) and interpret the result as a relation between the fine structure constant and the densities of various components of matter in the universe.

To conclude, we have considered the energy content of the Machian universe taking into account the existence of charged particles and their contribution to the total Machian energy. The energy of Machian interactions of all particles in the universe was identified with the dark energy. It has allowed us to derive a formula for the fine structure constant in terms of the radiation, baryon and dark energy densities. The formula agrees with the experimental value of the fine structure constant up to the uncertainty in the observed values of the cosmological parameters. Being depending on the energy densities of different components of matter in the universe, the formula might be helpful for understanding the recent evidence of the cosmological variations of the fine structure constant at high redshifts \cite{PRL}. The formula also yields a proportion between the densities of the baryon and dark energy components in the universe and a relation between the densities of various species of matter in the universe in terms on the observed (current) value of the fine structure constant. 


\section*{Acknowledgement:}

The work of M.G. was supported by Georgian National Science Foundation (grant GNSF-1-4/06). He also acknowledges ICTP (Trieste) for its hospitality. I.K. would like to thank KCIK in Sopot (Poland) for its kind hospitality.



\begin{thebibliography}{99}

\bibitem{Gog1} M. Gogberashvili,
              Eur. Phys. J. {\bf C 63} (2009) 317, 
              arXiv: 0807.2439 [gr-qc].

\bibitem{Gog2} M. Gogberashvili,
              Eur. Phys. J. {\bf C 54} (2008) 671, 
              arXiv: 0707.4308 [hep-th].

\bibitem{Gog3} M. Gogberashvili,
              arXiv: 0910.0169 [physics.gen-ph];
              arXiv: 1008.2544 [gr-qc].

\bibitem{Mach} J. Barbour and H. Pfister,
              {\it Mach's Principle: From Newton's Bucket to Quantum Gravity}
              (Birkh\"auser, Boston 1995) Einstein studies, Vol. 6; \\
               A. Ghosh,
              {\it Origin of Inertia} (Apeiron, Montreal 2000).

\bibitem{Exp} V.W. Hughes, H.G. Robinson and V. Beltran-Lopez,
              Phys. Rev. Lett. {\bf 4} (1960) 342; \\
              R.W.P. Drever,
              Phil. Mag. {\bf 6} (1961) 683.

\bibitem{large} P.A.M. Dirac,
               Proc. Roy. Soc. London {\bf A 338} (1974) 439.

\bibitem{La-Li} O. Lahav and A.R. Liddle,
               arXiv: 1002.3488 [astro-ph.CO].

\bibitem{PRL} J.K. Webb, et al.,
             arXiv: 1008.3907 [astro-ph.CO].

\end{thebibliography}
\end{document}